\documentclass[prb,preprint,aps,superscriptaddress,longbibliography]{revtex4-1}
\usepackage[]{color}
\usepackage{appendix}
\usepackage{graphicx, epstopdf}
\usepackage{subfigure}
\usepackage{float}
\usepackage{amsmath}
\usepackage{amsfonts}
\usepackage{amssymb}
\usepackage{bm}% bold math
\usepackage{subfigure}
\usepackage{caption}
\usepackage{hyperref}
\usepackage{xr}
\usepackage{lineno}
\usepackage{natbib}
\usepackage{soul}
\usepackage{amssymb}
\usepackage{multirow}
\usepackage[T1]{fontenc}
\usepackage{booktabs}
\usepackage{lipsum}
\usepackage{mathrsfs}
\usepackage{xr}
\begin{document}
\makeatletter
\newcommand{\rmnum}[1]{\romannumeral #1}
\newcommand{\Rmnum}[1]{\expandafter\@slowromancap\romannumeral #1@}
\makeatother

\title{NanoBTE: Fast Iterative Solution of the Phonon Boltzmann Transport Equation for Nanoscale Heat Transport}

\author{Hongjiang Chen}	
\thanks{These authors contributed equally to this work.}
\affiliation{College of Future Information Technology and Key Laboratory of Micro and Nano Photonic Structures (MOE), Fudan University, Shanghai 200433, China}

\author{Hai-Xuan Lin}	
\thanks{These authors contributed equally to this work.}
\affiliation{Department of Physics, Fudan University, Shanghai 200433, China}

\author{Xiaole Tian}	
\affiliation{College of Future Information Technology and Key Laboratory of Micro and Nano Photonic Structures (MOE), Fudan University, Shanghai 200433, China}

\author{Hongyu Chen}	
\affiliation{College of Future Information Technology and Key Laboratory of Micro and Nano Photonic Structures (MOE), Fudan University, Shanghai 200433, China}

\author{Juan Zhang}	
\affiliation{College of Future Information Technology and Key Laboratory of Micro and Nano Photonic Structures (MOE), Fudan University, Shanghai 200433, China}

\author{Shengyao Yan}	
\affiliation{College of Future Information Technology and Key Laboratory of Micro and Nano Photonic Structures (MOE), Fudan University, Shanghai 200433, China}

\author{Hezhu Shao}	
\affiliation{Wenzhou Key Laboratory of Micro-nano Optoelectronic Devices, College of Electrical and Electronic Engineering, Wenzhou University, Wenzhou 325035, China}

\author{Yu Wu}	
\affiliation{Advanced Thermal Management Technology and Functional Materials Laboratory, Ministry of Education Key Laboratory of NSLSCS, School of Energy and Mechanical Engineering, Nanjing Normal University, Nanjing 210023, China}

\author{Junjie Liu}	
\email{jj_liu@shu.edu.cn}
\affiliation{Department of Physics, Institute for Quantum Science and Technology, Shanghai Key Laboratory of High Temperature Superconductors, International Center of Quantum and Molecular Structures, Shanghai University, Shanghai, 200444, China}

\author{Heyuan Zhu}	
\affiliation{College of Future Information Technology and Key Laboratory of Micro and Nano Photonic Structures (MOE), Fudan University, Shanghai 200433, China}

\author{Hao Zhang}	
\email{zhangh@fudan.edu.cn}
\affiliation{College of Future Information Technology and Key Laboratory of Micro and Nano Photonic Structures (MOE), Fudan University, Shanghai 200433, China}
\affiliation{The State Key Laboratory of Photovoltaic Science and Technology, Fudan University, Shanghai 200433, China}

%\date{January 2026}

\begin{abstract}
	Nanoscale heat dissipation has become a critical challenge in advanced semiconductor devices, where phonon transport can strongly deviate from the classical Fourier description due to the boundary scattering and ballistic effects. In this work, we propose NanoBTE, a deterministic finite-volume solver for the non-gray phonon Boltzmann transport equation under the relaxation-time approximation. The solver supports complex two- and three-dimensional geometries, band-resolved phonon properties, discrete-ordinates angular quadrature, volumetric heat generation, and multiple phonon boundary conditions, including thermalizing, diffuse, and specular reflections. %To improve the efficiency of multiscale simulations, both sequential and synthetic iterative schemes are implemented, where the latter couples the microscopic phonon transport equation with a macroscopic diffusion-type temperature equation to accelerate convergence in near-diffusive regimes. 
    Both sequential and synthetic iterative options are implemented for the steady-state solution. Furthermore, NanoBTE adopts a band–direction task decomposition strategy, enabling efficient MPI-based CPU parallelization and GPU acceleration of the dominant sparse transport operations.
\end{abstract}

\flushbottom
\maketitle

\thispagestyle{empty}
\section{Introduction}

Traditional metal-oxide-semiconductor field-effect transistors (MOSFETs) have served as the foundation of modern integrated circuits for decades due to their simple planar structure and excellent manufacturability\cite{sze2006physics}. However, as device dimensions entered the deep-submicrometer and nanometer regimes, severe short-channel effects, including threshold voltage roll-off, drain-induced barrier lowering, and increased leakage currents, significantly degraded device performance\cite{ieee2022international,sze2006physics,tsividis2011operation}. To address these limitations, fin field-effect transistors (FinFETs) were introduced, employing a three-dimensional fin-shaped channel wrapped by the gate on multiple sides, which provided stronger electrostatic control over the channel, effectively suppressing leakage currents and extending Moore's law beyond the limits of planar MOSFETs\cite{duan2024mosfet,colinge2008finfets}. However, as the device size shrinks, quantum mechanical phenomena such as direct tunneling through ultrathin gate dielectrics and source-to-drain tunneling become increasingly significant, resulting in higher off-state leakage currents and reduced energy efficiency\cite{jelodar2015tunneling}. Manufacturing variability, process fluctuations, and reliability concerns also become more severe at nanometer dimensions\cite{duan2024mosfet}.

As device dimensions shrink and transistor density increases, power dissipation becomes concentrated within increasingly confined volumes. In FinFETs and particularly in GAAFETs, the nanoscale geometry and reduced phonon transport paths hinder efficient heat removal from the active region. The resulting temperature rise can degrade carrier mobility, reduce drive current, increase leakage power consumption, and accelerate aging mechanisms\cite{pop2010energy}. Elevated channel temperatures may also exacerbate reliability issues, including bias temperature instability (BTI), hot-carrier injection (HCI), and electromigration\cite{amrouch2019reliability}, ultimately shortening device lifetime\cite{wu2025self}. Moreover, the self-heating effects can lead to thermal coupling between neighboring transistors in highly integrated circuits\cite{liu2014ieee}, creating localized hotspots and further complicating thermal management\cite{pop2010energy}. Consequently, understanding and mitigating self-heating effects has become a critical research topic for maintaining performance, reliability, and energy efficiency in advanced nanoscale semiconductor devices.

At the nanoscale, phonon heat transport often deviates significantly from the classical Fourier description because the characteristic length of the device becomes comparable to the phonon mean free paths (MFPs)\cite{chen2001ballistic}, and the local equilibrium assumption due to the frequent diffusive scatterings required for Fourier's law no longer holds. Instead, in this regime, heat conduction becomes intrinsically non-local and involves the coexistence of diffusive, quasi-ballistic, and ballistic phonons\cite{chen2001ballistic,cahill2014nanoscale}.

This transport regime is commonly referred to as the Casimir-Knudsen regime. The Casimir aspect originates from boundary-limited phonon transport, where phonons are strongly scattered by device surfaces or interfaces and the effective phonon MFPs are limited by the characteristic device size. The Knudsen aspect emphasizes that similar non-local transport can also arise when the heat-source size itself becomes comparable to or smaller than the phonon MFPs, even in the absence of strong boundary scattering. 

Such non-Fourier phonon transport is particularly relevant to self-heating in aggressively scaled FinFETs and GAAFETs, where heat is generated in nanoscale channels and must be dissipated through confined geometries and heterogeneous interfaces. The phonon Boltzmann transport equation (BTE), provides a natural theoretical framework for describing heat conduction in the Casimir--Knudsen regime, because it explicitly accounts for phonon distributions, MFPs, group velocities, and boundary or interface scattering processes, although its numerical solution remains challenging. The distribution function depends not only on physical space, but also on phonon propagation direction, frequency or band index, and polarization, leading to a high-dimensional phase-space problem. Moreover, realistic nanoscale simulations require non-gray phonon properties, accurate treatment of phonon-boundary scattering, and stable coupling between the microscopic phonon distribution and macroscopic temperature fields.

Solving the phonon BTE has motivated extensive development of numerical methods for multiscale heat transport across the ballistic-diffusive regimes. Stochastic techniques, such as Monte Carlo (MC) methods,  provide high flexibility in handling complex geometries and full phonon dispersion, but their computational cost becomes prohibitive in the near-diffusive regime due to statistical noise and slow convergence. Conversely, deterministic approaches, including the discrete ordinate method (DOM), avoid statistical noise and offer higher accuracy.  However, they are bounded by high-dimensional discretization requirements and suffer from ray effects in multidimensional problems. To overcome these limitations, several hybrid and acceleration strategies have been proposed.  The discrete unified gas kinetic scheme (DUGKS)\cite{zhang2019discrete} achieves multiscale capability by coupling phonon advection and scattering processes at cell interfaces, ensuring robust performance across different transport regimes. To alleviate the convergence bottleneck of conventional DOM in the near-diffusive limits, macro-micro coupled acceleration strategies have been developed. These include unified implicit kinetic schemes (UIKS)\cite{zhang2017unified,zhang2019implicit}, which have been extended from gray transport approximations to  non-gray models by incorporating realistic phonon dispersion and polarization via specialized micro-inner iterations, and fast synthetic iterative schemes (FSIS)\cite{zhang2021fast}, which derive macroscopic moment or a synthetic diffusion equation from BTE. By tightly coupling these macroscopic fields with the microscopic kinetic solver, these methods facilitate efficient information exchange across scales and achieve orders-of-magnitude acceleration in multiscale heat transport simulations. Most recently, deep-learning-based paradigms, such as Monte Carlo physics-informed neural networks (MC-PINNs)\cite{lin2025monte}, have been proposed to fundamentally bypass the curse of dimensionality and grid dependencies, offering an alternative framework for high-dimensional BTE workflows.

It is worth mentioning that several general-purpose deterministic phonon BTE solvers have also been developed recently. GiftBTE \cite{hu2024giftbte,hu2022optimized,hu2024ultraefficient} employs a second-order finite-volume discrete-ordinates framework, together with sequential and synthetic iterative schemes, and supports complex geometries, volumetric heat sources, and multiple boundary conditions. JAX-BTE\cite{shang2025jax} introduced a GPU-accelerated and differentiable implementation based on the JAX framework, enabling both forward and inverse phonon transport simulations.

In this work, we present NanoBTE, a deterministic finite-volume solver for the steady-state, non-gray phonon BTE aimed at nanoscale heat transport in the Casimir-Knudsen regime. NanoBTE solver handles the complex two- and three-dimensional geometries of realistic semiconductor devices and resolves the phonon phase space through band-resolved properties and discrete-ordinates angular quadrature. Thermalizing, diffuse, and specular boundary conditions, together with volumetric heat generation, allow the surface scattering and localized self-heating of scaled transistors to be captured directly. To improve convergence across the ballistic-to-diffusive transition, both a sequential and a synthetic iterative scheme are available for solving the steady-state transport equation. To alleviate the high cost of deterministic BTE simulations, NanoBTE exploits the natural separability of the discretized equation in the band-direction phase space: independent transport tasks are distributed among MPI ranks for CPU parallelization, and their dominant sparse operations are offloaded to GPUs for further acceleration. The solver is validated on in-plane MoS$_2$ and cross-plane diamond benchmarks and on three-dimensional FinFET and GAAFET structures, demonstrating both the accuracy of the transport solution and the parallel efficiency of the implementation across a wide range of characteristic length scales.

\section{Methods}

A phonon represents a collective oscillation of atoms around their equilibrium lattice positions, rather than the motion of an individual atom. Each microscopic phonon mode is labeled by a wave vector $\mathbf k$ in the first Brillouin zone and a branch index $p$, with angular frequency $\omega_p(\mathbf k)$, group velocity $\mathbf v_p(\mathbf k)=\boldsymbol{\nabla}_{\mathbf k}\omega_p(\mathbf k)$, and modal energy $E_p(\mathbf k)=\hbar\omega_p(\mathbf k)$. The corresponding wavelength is defined from the magnitude of the wave vector as $\lambda_p(\mathbf k)={2\pi}/{|\mathbf k|}$, or, in an isotropic acoustic approximation, $\lambda_p\approx 2\pi v_p/\omega_p$.

In nanoscale and microscale systems where the characteristic length $L$ becomes comparable to, or smaller than the phonon mean free path $\Lambda$, strong localization is absent and coherent effects can be neglected. Phonon–phonon interactions are represented by a mode-dependent relaxation time $\tau$ that is independent of the non-equilibrium distribution, and each phonon mode is assumed to interact independently with system boundaries. Under these assumptions, phonon transport in this Casimir-Knudsen regime can be described by the particle-like Boltzmann transport equation within the relaxation-time approximation (RTA)\cite{ziman1960electrons,mazumder2021boltzmann,carslaw1959conduction}, 

\begin{equation}\label{eq: BTE-RTA}
\frac{\partial f_p}{\partial t}+\mathbf{v}_p\cdot \boldsymbol{\nabla }f_p=-\frac{f_p-f_{p}^{0}}{\tau}+s
\end{equation}

where $f_p=f_p\left( \omega ,\mathbf{k},\mathbf{r},t \right)$ denotes the local phonon distribution function, representing the average occupation of phonons with branch $p$, frequency $\omega$, and the wave vector $\mathbf{k}$ at position $\mathbf{r}$ and time $t$, and $\mathbf{v}_p=\boldsymbol{\nabla }_{\mathbf{k}}\omega _p\left( \mathbf{k} \right)$, and $s$ denotes an external source term. Since phonons are bosons, then the local equilibrium distribution $f_p^0$ can be expressed as the Bose-Einstein distribution
\begin{equation}
f_{p}^{0}=\frac{1}{\mathrm{e}^{\hbar \omega _p\left( \mathbf{k} \right) /k_{\mathrm{B}}T\left( \mathbf{r},t \right)}-1},
\end{equation}
where $\hbar$ is the reduced Planck constant, $k_{\mathrm{B}}$ is the Boltzmann constant, and $\omega_p(\mathbf{k})$ denotes the dispersion relation with branch $p$, $T(\mathbf{r},t)$ denotes the local temperature at position $\mathbf{r}$ and time $t$.
By defining the phonon energy density distribution $e_p=\hbar\omega_pD_p(\omega,\mathbf{k})f_p$, where $D_p(\omega,\mathbf{k})$ is the density of phonon states at $(p,\omega,\mathbf{k})$, Eq.~\eqref{eq: BTE-RTA} can be rewritten as,

\begin{equation}
    \frac{\partial e_p}{\partial t}+\mathbf{v}_p\cdot \boldsymbol{\nabla }e_p=-\frac{e_p-e_{p}^{0}}{\tau}+\dot{Q}.
    \label{eq:bte_eq}
\end{equation}

Here, a deterministic finite-volume framework called NanoBTE was developed for solving the steady-state and non-gray phonon BTE as described by Eq.~(\ref{eq:bte_eq}) in two-/three-dimensional geometries. The NanoBTE solver supports COMSOL and Gmsh meshes\cite{multiphysics1998introduction,geuzaine2009gmsh}, discrete ordinates angular quadrature, band-resolved phonon properties, volumetric heat sources, and thermalizing, diffuse, or specular boundary conditions\cite{zhang2021fast,hu2024giftbte}. As demonstrated in \ref{sec:bte}, a sequential iteration strategy was realized firstly, and subsequently a synthetic iterative scheme was implemented by coupling the BTE with a macroscopic diffusion-type temperature equation, which improves convergence in multiscale regimes while retaining the non-Fourier contribution from the phonon distribution.

As described in \ref{sec:sequential}, the sequential iterative strategy solves the phonon steady BTE ($\partial/\partial t=0$) and updates the lattice temperature through the local energy-conservation condition\cite{zhang2021fast,zhang2025synthetic}. Starting from a given lattice temperature field $T_L^{(n)}$, the local equilibrium distribution is constructed and the steady-state BTE for each discrete phonon mode is solved\cite{zhang2025synthetic},

\begin{equation}
    \mathbf{v}_{\lambda i}\cdot \boldsymbol{\nabla} e_{\lambda i}=-\frac{e_{\lambda i}-e_{\lambda}^{0}}{\tau_{\lambda}}+S_{\lambda i},
\end{equation}

where $\mathbf{v}_{\lambda i}=v_{\lambda}\mathbf{s}_i$ and $\mathbf{s}_i=\mathbf{k}_i/\left|\mathbf{k}_i\right|$, $\lambda$ labels the representative phonon band and $i$ labels the propagation direction after the band and direction discrete process. After obtaining the updated phonon distributions $e_{\lambda i}^{(n+1)}$, the lattice temperature is recomputed from the collision-term energy-conservation condition\cite{zhang2021fast},

\begin{equation}
    T_L^{(n+1)}=\frac{\sum_{\lambda}\frac{1}{\tau_{\lambda}}\sum_i w_i e_{\lambda i}^{(n+1)}}{\sum_{\lambda}\frac{C_{\lambda}}{\tau_{\lambda}}} .
\end{equation}

Thus, one sequential iteration can be summarized as\cite{zhang2021fast}

\begin{equation}
    T_L^{(n)}\rightarrow e_{\lambda i}^{(n+1)}\rightarrow T_L^{(n+1)} .
\end{equation}

The macroscopic temperature and heat flux are then evaluated from the moments of the phonon distribution. This procedure is repeated until the prescribed convergence criteria are satisfied. The sequential scheme is simple and robust, but its convergence may become slow in near-diffusive regimes because the temperature field is updated only through local moment relations.

As described in \ref{sec:synthetic}, the synthetic iterative algorithm accelerates the convergence of the NanoBTE solver by introducing an additional macroscopic diffusion equation for the lattice temperature. Different from the sequential scheme, where \(T_L\) is updated only from the local collision-moment relation, the synthetic scheme updates \(T_L\) through a macro-micro coupled procedure. After solving the microscopic transport equation for all discrete band-direction modes, the heat flux is reconstructed from the non-equilibrium phonon energy distribution as\cite{zhang2025synthetic}
\begin{equation}
\mathbf{q}^{(n+1)}=\sum_{\lambda}\sum_i w_i \mathbf{v}_{\lambda i}e_{\lambda i}^{(n+1)},
\label{eq:discrete_heat_flux}
\end{equation}

where \(w_i\) is the angular quadrature weight, \(\lambda\) denotes the representative phonon band, and \(i\) denotes the discrete propagation direction.

In the diffusive limit, the heat flux should approach the Fourier contribution \(-\kappa_{\mathrm{bulk}}\nabla T_L\). Therefore, the total heat flux can be decomposed into a Fourier-like part and a non-Fourier correction\cite{zhang2021fast,zhang2025synthetic} as

\begin{equation}
\mathbf{q} = -\kappa_{\mathrm{bulk}}\boldsymbol{\nabla} T_L + \mathbf{q}_{\mathrm{nonF}} ,
\label{eq:flux_decomposition}
\end{equation}
where \(\kappa_{\mathrm{bulk}}\) is the bulk thermal conductivity and \(\mathbf{q}_{\mathrm{nonF}}\) represents the kinetic correction extracted from the nonequilibrium phonon distribution. In the iterative implementation, this correction is evaluated as\cite{zhang2021fast,hu2024giftbte}

\begin{equation}
\mathbf{q}_{\mathrm{nonF}}^{(n+1)} =\mathbf{q}^{(n+1)}+\kappa_{\mathrm{bulk}}\boldsymbol{\nabla} T_L^{(n)} .
\label{eq:q_nonF}
\end{equation}
This term contains the ballistic and non-local components of heat transport that cannot be represented by the local Fourier temperature gradient.

Substituting Eq.~\eqref{eq:flux_decomposition} into the steady energy-conservation equation gives the synthetic macroscopic diffusion equation\cite{zhang2021fast},

\begin{equation}
\boldsymbol{\nabla}\cdot\left(-\kappa_{\mathrm{bulk}}\boldsymbol{\nabla}T_L^{(n+1)}\right)=\dot{Q}-\boldsymbol{\nabla}\cdot \mathbf{q}_{\mathrm{nonF}}^{(n+1)} ,
\label{eq:synthetic_diffusion}
\end{equation}

where \(\dot{Q}\) is the volumetric heat-generation rate. The updated lattice temperature is then used to reconstruct the local equilibrium energy distribution\cite{zhang2021fast},

\begin{equation}
e_{\lambda}^{0,(n+1)}=\frac{C_{\lambda}}{W_{\Omega}}T_L^{(n+1)},\qquad
W_{\Omega}=\sum_i w_i ,
\label{eq:equilibrium_energy_update}
\end{equation}

which provides the equilibrium state for the next microscopic BTE solve.

Thus, one synthetic iteration can be written as\cite{zhang2021fast}
 \begin{equation}
     % \mathbf{q}=-\kappa_{\mathrm{bulk}}\boldsymbol{\nabla} T_L+\mathbf{q}_{\mathrm{nonF}},
     T_L^{(n)}\rightarrow e_{\lambda i}^{(n+1)}\rightarrow\mathbf{q}_{\mathrm{nonF}}^{(n+1)}\rightarrow T_L^{(n+1)} ,
 \end{equation}

Compared with the sequential update, this synthetic procedure propagates long-range temperature information through a global diffusion-type equation, rather than through purely local moment updates. As a result, the slowly decaying diffusive error modes are damped more efficiently, which improves the convergence of deterministic BTE simulations in the near-diffusive regime while retaining the non-Fourier kinetic correction obtained from the phonon distribution.

\begin{figure}[ht!]
    \centering
    \includegraphics[width=0.9\linewidth]{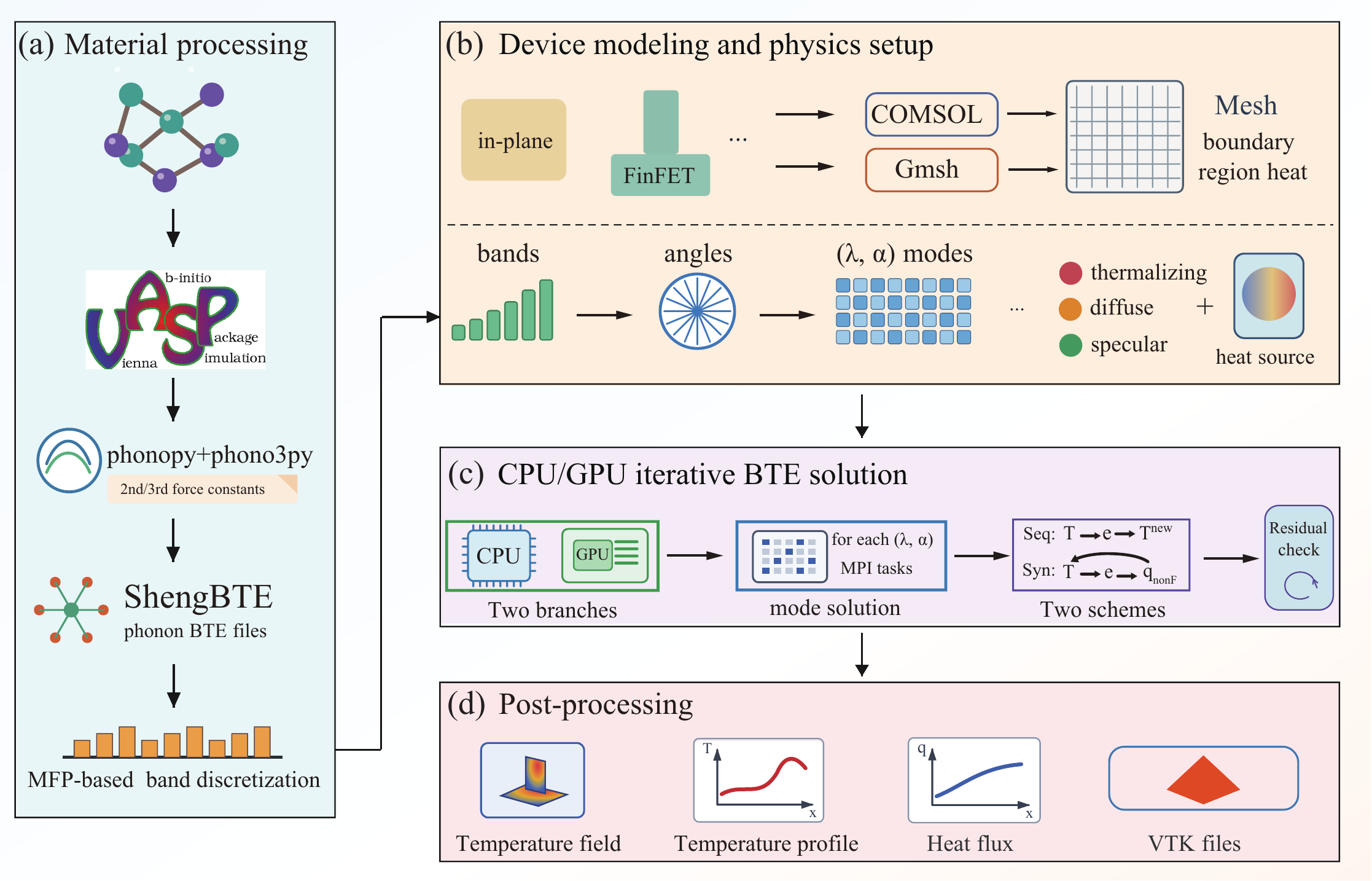}
    \caption{Workflow of NanoBTE. (a) Material processing, where crystal structures are converted into band-resolved phonon properties through first-principles calculations, phonon-property calculations, and mean-free-path-based band discretization. (b) Device modeling and physics setup, including geometry and mesh generation, boundary-condition assignment, heat-source definition, and band–direction mode construction. (c) CPU- or GPU-accelerated iterative solution of the phonon Boltzmann transport equation using sequential or synthetic schemes. (d) Post-processing of temperature fields, temperature profiles, heat-flux distributions, and visualization files.}
    \label{fig:workflow}
\end{figure}

The complete workflow of NanoBTE is summarized in Figure~\ref{fig:workflow}. The workflow connects material-level phonon property preparation with device-level BTE simulation. In the material-level part, \textit{ab-initio} calculations are used to obtain the second- and third-order force constants of device materials, which are then used as inputs for phonon BTE solver for materials such as ShengBTE\cite{li2014shengbte} to generate the phonon properties required by the NanoBTE solver. In the device simulation part, the device geometry is constructed and spatially discretized, with the computational mesh generated with boundary labels and heat-source regions. The angular quadrature is applied to discretize the phonon propagation directions.
After the preparation procedures, the phonon BTE equations are solved based on the finite-volume algorithm, implemented by using both the CPU-and GPU-acceleration methods. Both implementations follow the same physical model and iterative framework. Finally, the converged phonon distribution is post-processed to obtain macroscopic quantities, including the temperature field, temperature profiles, heat-flux profiles, and visualization files.

In the NanoBTE solver, both CPU- and GPU-based acceleration are realized for the sequential and synthetic iterative algorithms, following a common design principle: precompute what is iteration-independent, parallelize over what is mode-independent, and offload what is data-parallel. As detailed in \ref{sec:parallel}, after angular and spectral discretization each pair $(\lambda,i)$ leads to an independent sparse transport system of the form $(I/\tau_\lambda+v_\lambda K_i)\Delta e_{\lambda i}=R_{\lambda i}$, as described in Eq.~(\ref{eq:mpi}), where the unit advection operator $K_i$ depends only on the mesh and the propagation direction. Since the transport matrix does not change between outer iterations, it is factorized once before the iteration loop,
\begin{equation}
A_{\lambda i}=\frac{1}{\tau_{\lambda}}\mathbf{I}+v_{\lambda}K_{i}
\;\xrightarrow{\ \text{once}\ }\;
L_{\lambda i}U_{\lambda i},
\qquad (\lambda,i)\in\mathcal{T}_{r},
\label{eq:lu_cache}
\end{equation}
so that each subsequent iteration reduces to right-hand-side assembly and inexpensive triangular substitutions. The same precompute-and-cache principle is applied to the geometry: cell--face connectivity, direction-resolved advection coefficients, gradient and vertex-interpolation operators, and the boundary caches are all constructed once from the mesh.

Because the $(\lambda,i)$ transport tasks are mutually independent within one iteration and are coupled only through the macroscopic fields, they are distributed among MPI ranks in a round-robin manner, while the mesh and the phonon distribution arrays are replicated on every rank. Each rank assembles and factorizes only its own subset of transport systems, so both the factorization time and the LU memory footprint decrease as ranks are added. One accelerated iteration can thus be summarized compactly as
\begin{equation}
e^{(n+1)}=e^{(n)}
+\underbrace{\sum_{r=1}^{N_{\mathrm{rank}}}\;\sum_{(\lambda,i)\in\mathcal{T}_{r}}
\bigl(L_{\lambda i}U_{\lambda i}\bigr)^{-1}R^{(n)}_{\lambda i}}_{\text{CPU: cached LU solves}\;+\;\text{\texttt{Allreduce}}},
\qquad
\bigl\{T,\;\mathbf{q},\;\boldsymbol{\nabla}\!\cdot\!\mathbf{q}_{\mathrm{nonF}}\bigr\}
=\underbrace{\mathcal{M}\bigl[e^{(n+1)}\bigr]}_{\text{GPU: chunked reductions}},
\label{eq:accel_split}
\end{equation}
where $\mathcal{T}_{r}$ denotes the round-robin subset of band--direction tasks assigned to rank $r$ out of $N_{\mathrm{rank}}$ ranks, each rank computes only its own slices of the update tensor and the outer sum is realized by a single MPI \texttt{Allreduce}, and $\mathcal{M}$ collects the phase-space reductions of Eq.~\eqref{eq:discrete_heat_flux} and the face-based divergence of Eq.~\eqref{eq:q_nonF}. The lattice temperature $T_L^{(n+1)}$ then follows either from the collision moment (sequential scheme) or from the cached macroscopic LU solve of Eq.~\eqref{eq:synthetic_diffusion} (synthetic scheme). This task-parallel organization avoids spatial-domain decomposition altogether, and thereby keeps unstructured meshes, complex boundary conditions, and global temperature updates as simple as in the serial code.

The reduction operator $\mathcal{M}$ in Eq.~\eqref{eq:accel_split} comprises the operations that remain the per-iteration bottleneck once the factorizations are cached, namely the moment evaluations of $T_L$, $T$, and $\mathbf{q}$, and, in the synthetic scheme, the face-based assembly of $\boldsymbol{\nabla}\cdot\mathbf{q}_{\mathrm{nonF}}$. These operations are data-parallel over cells, directions, and bands and are therefore offloaded to GPUs: geometric and spectral constants are uploaded to device memory once, the distribution tensor is streamed through the device in cell-wise chunks to bound the memory footprint when several MPI ranks share one GPU, and only the resulting cell-wise fields are returned to the host. The accuracy-critical sparse LU solves deliberately remain on the CPU, and the GPU path falls back to the CPU implementation whenever a device or kernel is unavailable, so that GPU acceleration acts as an optional performance layer rather than a correctness requirement.

\section{Numerical results}
\subsection{Two-dimensional in-plane molybdenum disulfide and diamond cross-plane }

\begin{figure}[ht!]
    \centering
    \includegraphics[width=1\linewidth]{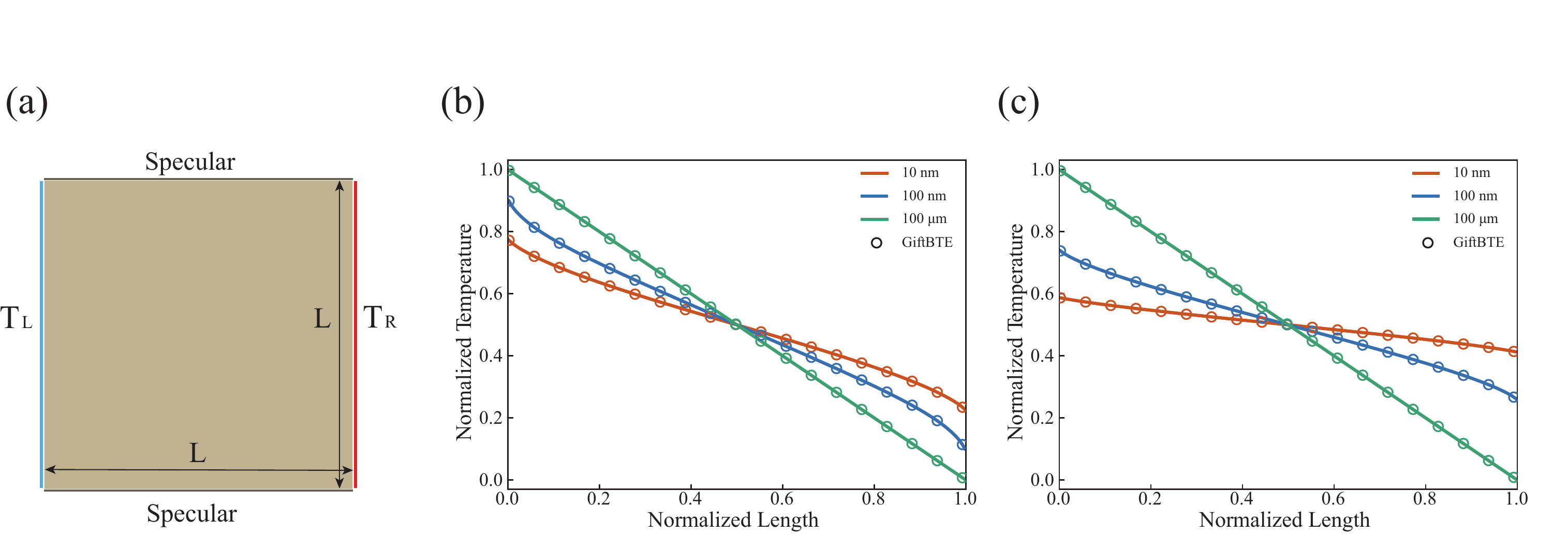}
    \caption{Comparison of in-plane thermal transport in MoS$_2$ and cross-plane thermal transport in diamond calculated using NanoBTE and GiftBTE\cite{hu2024giftbte,hu2022optimized,hu2024ultraefficient}. (a) Schematic of the square MoS$_2$ computational domain and boundary condition setup. (b) Dimensionless temperature profiles for in-plane thermal transport in MoS$_2$ at different characteristic lengths, $L=10 ~\mathrm{nm}$, $L=100 ~\mathrm{nm}$, and $L=100 ~\mu\mathrm{m}$. (c) Dimensionless temperature profiles for cross-plane thermal transport in diamond at different characteristic lengths, $L=10 ~\mathrm{nm}$, $L=100 ~\mathrm{nm}$, and $L=100 ~\mu\mathrm{m}$.} 
    \label{fig:MoS2_structure_temperature}
\end{figure}

The in-plane thermal transport in monolayer molybdenum disulfide (MoS$_2$) and the diamond cross-plane are first investigated as benchmark problems to validate the capability of the present NanoBTE solver for size-dependent phonon transport. In particular, the MoS$_2$ case represents in-plane transport in a two-dimensional material, whereas the diamond case represents cross-plane transport in a three-dimensional crystal. By comparing different characteristic lengths, the transition from ballistic or quasi-ballistic transport to the nearly diffusive regime can be clearly captured.
As shown in Figure~\ref{fig:MoS2_structure_temperature}(a), the computational domain is a square MoS$_2$ region with side length $L$. A reference temperature of $T_0=300~\mathrm{K}$ is used in this calculation. To impose a temperature bias around the reference state, the left and right boundaries are set as the thermalizing boundaries with fixed temperature of $T_L=305~\mathrm{K}$ and $T_R=295~\mathrm{K}$. These two boundaries act as external thermal reservoirs. The upper and lower boundaries are treated as specularly reflecting boundaries. Therefore, no artificial energy loss is introduced through the top and bottom sides. Heat transport is driven by the temperature difference along the in-plane direction, while the transverse direction remains adiabatic.

For the numerical setup, the computational domain is discretized by a structured quadriliteral mesh with $200\times200$ cells, corresponding to 40,000 elements in total. The phonon dispersion is discretized into 12 bands, and the angular space is discretized into 32 directions in total. The phonon properties of MoS$_2$ are obtained from the first-principles phonon calculations by using Phonopy and ShengBTE\cite{togo2023first,li2014shengbte}. Accumulant lattice thermal conductivity results is provided in Figure~\ref{fig:fig1}, the MFPs contributing 50\% to the total $\kappa_L$ are $\sim$150-200 nm for MoS$_2$, $\sim$200 nm for Si and $\sim$250-400 nm for diamond. Thus, three characteristic lengths are considered, namely $L=10 ~\mathrm{nm}$(ballistic), $L=100 ~\mathrm{nm}$(quasi-ballistic), and $L=100 ~\mu\mathrm{m}$(diffusive)\cite{yang2015heating}. These three cases are designed to cover different phonon transport regimes. 
For comparison among different scales, the temperature $T$ and spatial coordinate $X$ are normalized as $T^* = \frac{T - T_R}{T_L - T_R}$ and $X^* = \frac{x}{L}$, respectively.

Figure~\ref{fig:MoS2_structure_temperature}(b) shows the dimensionless temperature profiles along the in-plane transport direction. For the $L=100~\mu\mathrm{m}$ case, the temperature distribution is nearly linear from the hot side to the cold side. In this case, the phonon transport is close to the diffusive regime. Consequently, a well-defined temperature gradient is established, whereby the BTE solution asymptotes to the classical Fourier-like linear temperature profile. Specifically, the 10~nm case produces obvious temperature jumps near both thermalizing boundaries. Meanwhile, the temperature variation in the interior region becomes much flatter than the diffusive linear profile. This behavior is a typical signature of quasi-ballistic phonon transport\cite{chen2021non}. Since the system length is very small, the phonon distribution inside the domain is therefore strongly non-equilibrium, and the boundary thermalization process leads to an apparent temperature slip near the contacts.
The 100 nm case exhibits an intermediate behavior between the 10 nm and 100 $\mu$m cases. This trend indicates that as the characteristic length $L$ increases, the Knudsen number $K_n=\Lambda/L$ gradually decreases, and the phonon scattering inside the system becomes more complete. Therefore, the heat transport process gradually transitions from ballistic dominated to quasi-diffusive transport, and eventually approaches the diffusive transport region.

Figure~\ref{fig:MoS2_structure_temperature}(c) shows the calculated normalized cross-plane temperature profiles of the diamond cross-plane for different characteristic lengths. The computational domain, mesh resolution, boundary-condition setup, and characteristic lengths for diamond cross-plane are kept the same as those in the MoS$_2$ case. In the input setting, diamond is treated as a three-dimensional material and its phonon properties are represented by 12 effective bands.
Similar to the MoS$_2$ case, the smallest diamond film exhibits pronounced temperature jumps near the thermalizing contacts, reflecting the strong influence of boundary thermalization and ballistic phonon transport. As the film thickness increases, the boundary temperature jumps are gradually reduced, and the temperature distribution becomes closer to a linear profile. This trend confirms that the diamond cross-plane system also undergoes a ballistic-to-diffusive transition as the characteristic length increases.

\subsection{Three-dimensional Field-effect-transistors Devices}

\subsubsection{Fin Field-effect Transistors}

To evaluate the capability of the present NanoBTE solver for nanoscale thermal transport, a fin field-effect transistor (FinFET) is selected as the benchmark case\cite{hisamoto2000finfet}. This architecture provides a representative transistor geometry where the characteristic device length scale approaches or falls below the phonon mean free path $\Lambda$.

\begin{figure}[ht!]
    \centering
    \includegraphics[width=0.7\linewidth]{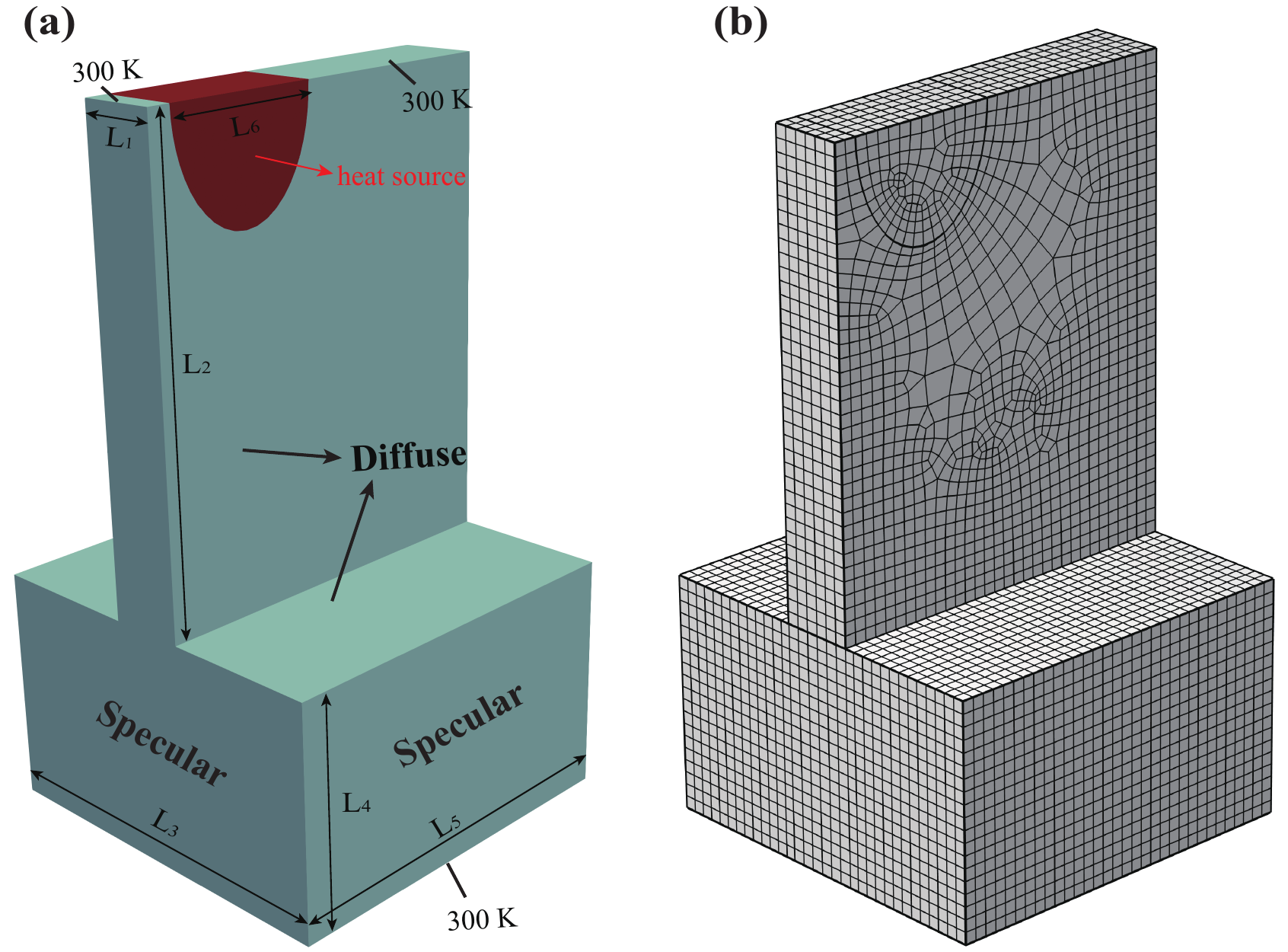}
    \caption{FinFET structure and mesh used for the three-dimensional NanoBTE calculation. (a) Schematic of the FinFET structure, including the substrate and fin body, together with the boundary-condition setup and heat-source assignment. (b) Mesh generated for the FinFET computational domain.}
    \label{fig:finfet_structure}
\end{figure}

As shown in Figure~\ref{fig:finfet_structure}(a), the computational domain consists of a silicon substrate, a vertical fin body, and a localized hot-spot region embedded near the top of the fin. The hot spot is modeled as an elliptical-cylinder region along the fin-width direction, which is used to simulate the Joule self-heating effects and electron-phonon energy relaxation localized near the drain side, where the electric field and carrier energy dissipation are usually the strongest. In actual transistor operation, the dissipation of electric field and carrier energy is usually concentrated near the drain and electrons transfer energy to the lattice through electron phonon scattering. Therefore, heat is often manifested as local hot-spot.

For simplification, in the BTE calculation, thermalizing boundaries at 300 K are set on the substrate bottom surface and on top external surfaces, representing thermal contact with external heat reservoirs or metal electrodes. The exposed side surfaces of the fin are assigned diffusely reflecting boundary conditions. This is because in practical nano-devices, silicon surfaces often contact with oxide layers, dielectric layers or rough interfaces. For nano-scale FinFET, this surface scattering significantly reduces the effective thermal conductivity, making it more difficult for heat to diffuse out of the hot-spot area. The boundary on the lateral side of substrate is set as a specular reflection boundary to simulate the symmetrical boundary in the device array. This processing can avoid unrealistic heat losses caused by the side boundaries of limited computational area. 

The characteristic sizes of the FinFET structure are labeled as $L_{1}$-$L_{6}$ in Figure~\ref{fig:finfet_structure}. Three geometrically scaled cases are considered, corresponding to fin widths of 10 nm, 100 nm, and 1 $\mu$m. All geometric dimensions are scaled proportionally, as summarized in Table~\ref{tab:finfet_dimensions}. These three cases are designed to cover different transport regimes, ranging from strongly size-dependent ballistic transport at the 10 nm scale to nearly diffusive transport at the 1 $\mu$m scale. These FinFET geometries are designed to span different Knudsen-number regimes. As the fin width decreases from 1 $\mu$m to 10 nm, the characteristic length becomes increasingly comparable to the phonon mean free paths in silicon, and boundary scattering as well as quasi-ballistic transport become more pronounced.

\setlength{\tabcolsep}{25pt}
\begin{table}[ht!]
    \centering
    \caption{Characteristic sizes of the FinFET structures.}
    \label{tab:finfet_dimensions}
    \begin{tabular}{cccc}
        \hline
        Location & 10 nm case & 100 nm case & 1000 nm case \\
        \hline
        $L_1$ & 10 & 100 & 1000 \\
        $L_2$ & 60 & 600 & 6000 \\
        $L_3$ & 50 & 500 & 5000 \\
        $L_4$ & 30 & 300 & 3000 \\
        $L_5$ & 50 & 500 & 5000 \\
        $L_6$ & 20 & 200 & 2000 \\
        \hline
    \end{tabular}
\end{table}

The heat-source intensity is also scaled with the device size. The volumetric heat-generation rates are set to $1.5\times10^{19}$, $1.5\times10^{17}$, and $1.5\times10^{15}~\mathrm{W/m^3}$ for the 10 nm, 100 nm, and 1 $\mu$m cases, respectively. The reason for this scaling is that, in the classical heat-diffusion limit, the characteristic temperature rise induced by a volumetric heat source approximately follows,

\begin{equation}
    \Delta T \sim \frac{Q L^2}{\kappa}
\end{equation}

where $Q$ is the volumetric heat-generation rate, $L$ is the characteristic length, and $\kappa$ is the bulk thermal conductivity. Therefore, when the device size is increased by a factor of 10, reducing the heat-generation rate by a factor of $10^2$ keeps the Fourier-limit temperature rise at a comparable level. 

\begin{figure}[ht!]
    \centering
    \includegraphics[width=0.8\linewidth]{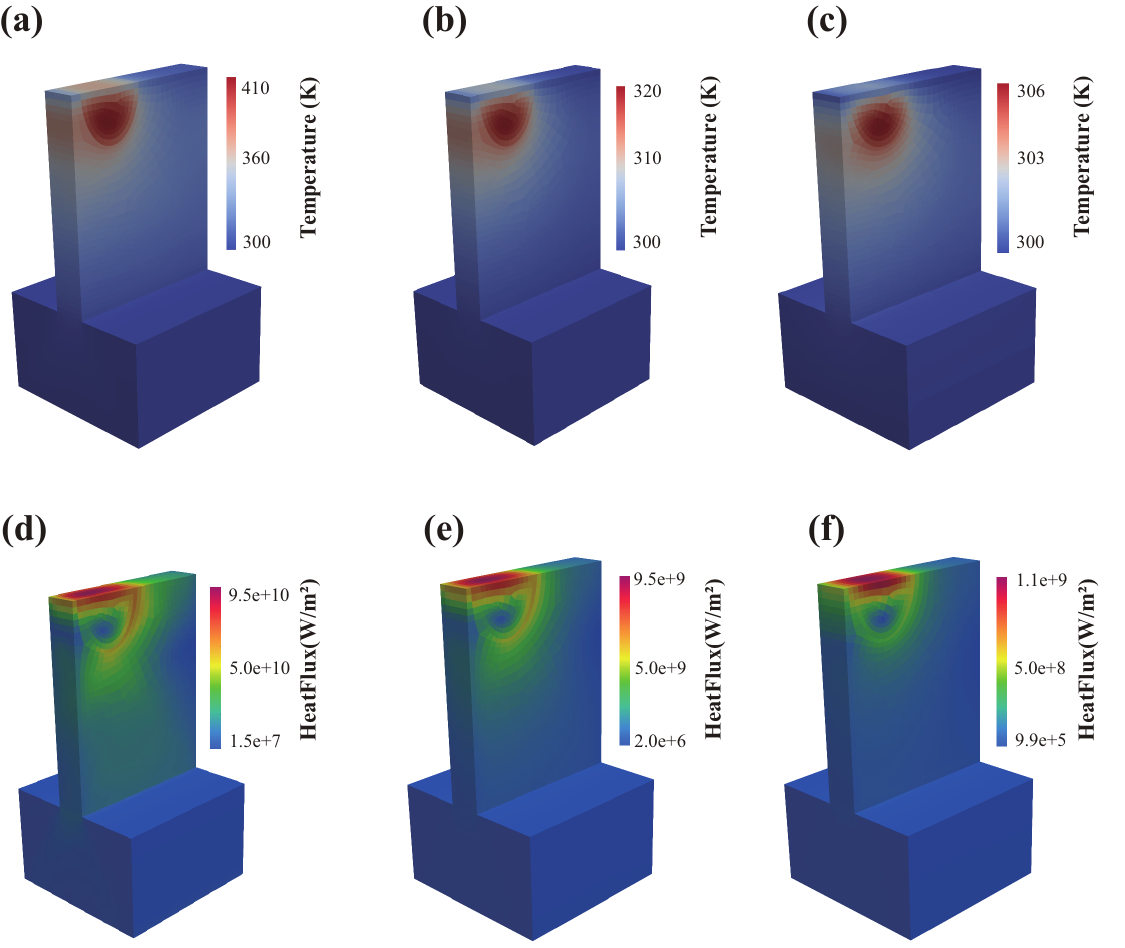}
    \caption{Numerical results of the FinFET cases with different fin widths obtained from the NanoBTE. (a) Temperature distribution for the 10 nm case. (b) Temperature distribution for the 100 nm case. (c) Temperature distribution for the 1 $\mu$m case. (d) Heat-flux magnitude distribution for the 10 nm case. (e) Heat-flux magnitude distribution for the 100 nm case. (f) Heat-flux magnitude distribution for the 1$\mu$m case.}
    \label{fig:finfet_result}
\end{figure}

In the NanoBTE solver, the heat source is treated as an equilibrium phonon source. It is uniformly distributed over all propagation directions and assigned to different phonon bands according to their modal heat capacities. This treatment assumes that the energy released from the electronic system is converted into lattice phonon energy locally. 
Based on the convergence test, for silicon, we classify 15 phonon bands, 64 directions, and 25,840 meshes for our analysis. Figure~\ref{fig:finfet_result} shows the temperature and heat-flux distributions of the FinFET structure with fin width of 10nm, 100nm and 1$\mu$m. 

\begin{figure}[ht!]
    \centering
    \includegraphics[width=1\linewidth]{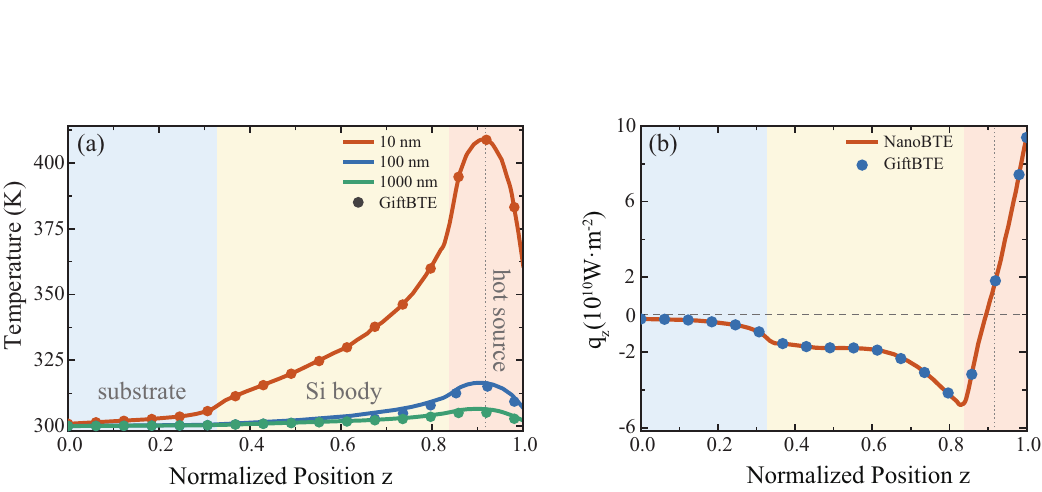}
    \caption{Line-profile comparison of the FinFET simulations obtained from NanoBTE and GiftBTE\cite{hu2024giftbte,hu2022optimized,hu2024ultraefficient}. (a) Temperature profiles extracted along the normalized vertical position for the FinFET structures with fin widths of 10 nm, 100nm, and 1000nm. (b) Distribution of the vertical heat-flux component along the same normalized path for the 10nm FinFET case.}
    \label{fig:finfet_result2}
\end{figure}

The corresponding temperature and heat-flux line-profile are presented in Figure~\ref{fig:finfet_result2}. A clear size dependence is observed in the temperature field. In the 10 nm case, the maximum temperature reaches approximately 410 K, corresponding to a temperature rise of about 110 K above the ambient temperature. When the fin width is increased to 100 nm, the peak temperature decreases to about 320 K. For the 1 $\mu$m case, the peak temperature is only about 306 K, and the temperature field becomes much smoother. Since the heat-source intensities have been scaled to give comparable Fourier-limit temperature rises, this reduction in temperature rise cannot be attributed simply to a lower heating power. Instead, it indicates the transition of the thermal transport regime. At the 10 nm scale, boundary scattering and ballistic phonon transport strongly suppress the effective thermal conductivity, producing a larger local thermal resistance and a stronger hot spot. As the characteristic length increases, phonon transport becomes more diffusive, and heat can spread more efficiently through the fin and substrate.

\subsubsection{Gate-all-around Field-effect Transistors}

\begin{figure}[ht!]
    \centering
    \includegraphics[width=0.9\linewidth]{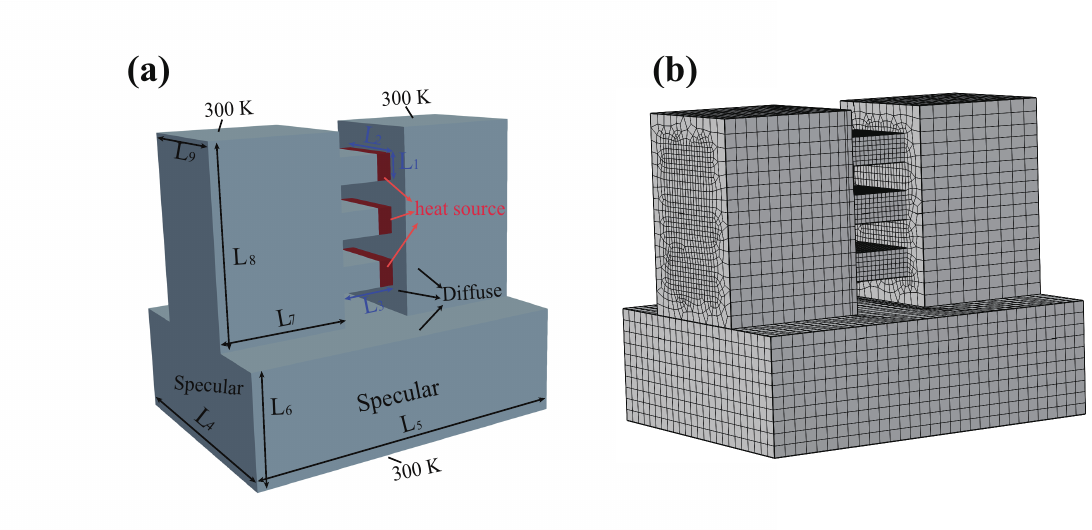}
    \caption{GAA structure and mesh used for the NanoBTE calculation. (a) Schematic of the three-stacked nanosheet gate-all-around FET (NS-GAAFET) structure, together with the boundary-condition setup and heat-source assignment. (b) Mesh generated for the GAA computational domain.}
    \label{fig:gaa_structure}
\end{figure}

To further validate the NanoBTE solver's applicability to more advanced transistor architectures, we then select a three-stacked nanosheet gate-all-around field-effect transistor (NS-GAAFET)\cite{loubet2017stacked}. Compared with the FinFET, as shown in Figure~\ref{fig:gaa_structure}, the GAAFET has a more complex geometry, including multiple vertically stacked nanosheet channels and source/drain blocks. Each nanosheet layer may generate localized heat, and thermal coupling occurs between different channels through the surrounding silicon regions. Thus, this case better tests the solver's capability in handling complex geometries and multi-hotspot coupled thermal transport problems.

As shown in Figure~\ref{fig:gaa_structure}(a), the computational domain consists of a silicon substrate, source and drain blocks, and a three-layer stacked nanosheet channel. For each nanosheet, a separate drain-side hot source region is defined. The thermal source configuration is also derived from the physical process of transistor operation: during device conduction, the electric field near the drain side is strong, leading to significant carrier energy dissipation. Electrons transfer energy to the lattice through electron-phonon interactions, resulting in local heat source near the drain side of each nanosheet channel. Since the GAAFET features three layers of channels, the thermal sources are locally distributed hotspots aligned vertically.

The boundary conditions are chosen similarly to those in the FinFET case. Thermalizing boundaries at 300 K are applied to the bottom surface of the substrate and to the source/drain contact surfaces. These boundaries represent thermal contact with metal electrodes or external heat sinks and allow phonons to be absorbed and re-emitted in equilibrium at the ambient temperature. The exposed nanosheet surfaces and other non-contact silicon surfaces are treated as diffusely reflecting boundaries, which account for phonon scattering at rough free surfaces or dielectric interfaces surrounding the nanosheets. The far-field side surfaces of the substrate are modeled as specularly reflecting boundaries to avoid artificial lateral heat loss through the truncated computational domain. 

\begin{figure}[ht!]
    \centering
    \includegraphics[width=0.8\linewidth]{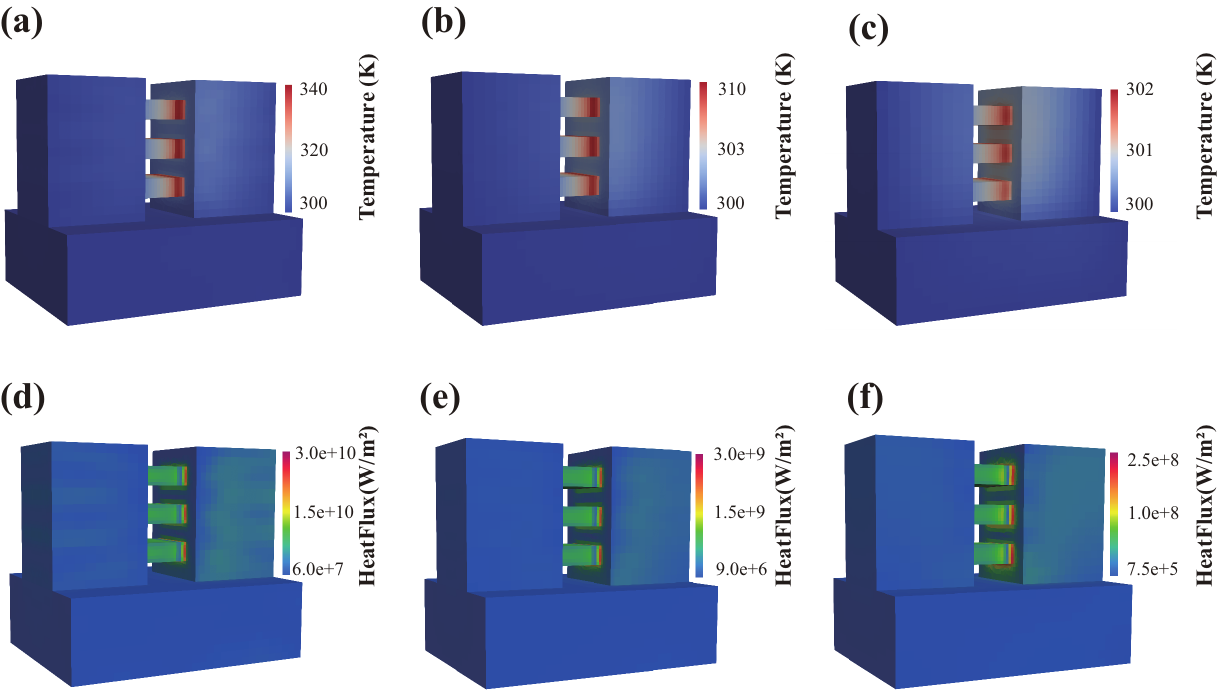}
    \caption{Temperature and heat-flux magnitude distributions of the GAAFET structures with different nanosheet heights obtained from the NanoBTE. (a) Temperature distributions for the 5 nm case. (b) Temperature distributions for the 50 nm case. (c) Temperature distributions for the 500 nm case. (d) Heat-flux magnitude distribution for the 5 nm case. (e) Heat-flux magnitude distribution for the 50 nm case. (f) Heat-flux magnitude distribution for the 500nm case.}
    \label{fig:gaa_result}
\end{figure}

The characteristic sizes labeled in Figure~\ref{fig:gaa_structure} are summarized in Table~\ref{tab:gaa_dimensions_5nm}. The values listed in the table correspond to the 5 nm case, where the nanosheet height is 5 nm. The 50 nm and 500 nm cases are obtained by increasing all geometric sizes by factors of 10 and 100, respectively. Therefore, the three GAAFET cases provide a scaled comparison from a strongly confined nanosheet structure to a larger structure approaching the diffusive regime.

\begin{table}[ht!]
    \centering
    \caption{Structural parameters of the GAAFET structure for the 5 nm case.}
    \label{tab:gaa_dimensions_5nm}
    \setlength{\tabcolsep}{10pt}
    \begin{tabular}{cccccccccc}
        \hline
        Location & $L_1$ & $L_2$ & $L_3$ & $L_4$ & $L_5$ & $L_6$ & $L_7$ & $L_8$ & $L_9$ \\
        \hline
        Length (nm) & 5 & 15 & 3 & 50 & 64 & 20 & 14 & 35 & 25 \\
        \hline
    \end{tabular}
\end{table}

The volumetric heat-generation rates assigned to the hot-spot regions are $1.5\times10^{19}$, $1.5\times10^{17}$, $1.5\times10^{15}~\mathrm{W/m^3}$ for the 5 nm, 50 nm, and 500 nm cases, respectively, similar to those in the FinFET case. Since device sizes are increased by a factor of 10 between adjacent cases, the heat-generation rate is reduced by a factor of $10^2$ to keep the corresponding Fourier-limit temperature rise comparable. This allows the BTE solver's results to highlight the influence of phonon size effects and boundary scattering, rather than being dominated by the change in absolute heating intensity. 

The same phonon and direction discretization as used in the FinFET simulations is adopted for the GAAFET cases. Based on the convergence test, the computational domain is discretized using a mesh with 30,173 volume elements. The calculated temperature and heat-flux distributions of the three GAAFET cases with nanosheet heights of 5 nm, 50 nm, and 500 nm are shown in Figure~\ref{fig:gaa_result}.

\begin{figure}[ht!]
    \centering
    \includegraphics[width=1\linewidth]{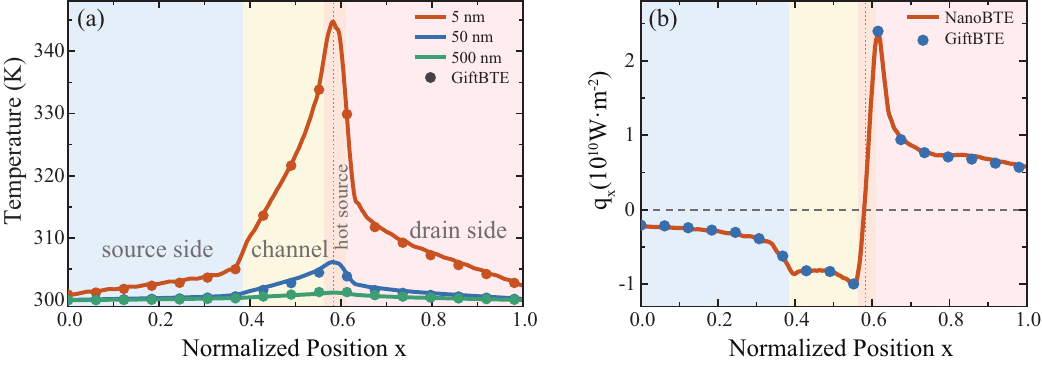}
    \caption{Line-profile comparison of the GAAFET simulations obtained from NanoBTE and GiftBTE\cite{hu2024giftbte,hu2022optimized,hu2024ultraefficient}. (a) Temperature profiles extracted along the normalized direction of a selected silicon nanosheet channel for the GAAFET structures with nanosheet heights of 5 nm, 50 nm, and 500 nm. (b) Distribution of the heat-flux component along the same normalized direction of the selected silicon nanosheet channel for the 5 nm case.}
    \label{fig:gaa_result2}
\end{figure}

%The size dependence is again evident. 
The corresponding temperature and heat-flux line-profile are presented in Figure~\ref{fig:gaa_result2}. For the 5 nm case, the peak temperature is approximately 340 K, indicating a significant local temperature rise caused by the strong confinement of phonon transport in the nanosheet structure. For the 50 nm case, the peak temperature decreases to about 310 K. For the 500 nm case, the maximum temperature is approximately 302 K. This trend indicates that as the nanosheet dimension increases, phonon transport gradually approaches the diffusive limit and thermal resistance associated with boundary scattering is reduced. 

Compared with the FinFET, the GAAFET exhibits a more distributed and coupled heating behavior. In the FinFET, the heat source is mainly localized near the top region of a single fin. In contrast, the GAAFET contains three vertically stacked nanosheet channels, and each channel has its own drain-side heat source. Therefore, the resulting temperature field is formed by the coupling of multiple hot spots. Heat generated in each nanosheet can spread not only along the channel direction, but also through the source/drain blocks and the substrate. This additional three-dimensional heat-spreading path makes the GAAFET thermal field more complex than FinFET. Nevertheless, both devices show the same fundamental trend: reducing the characteristic sizes enhances boundary scattering, suppresses the effective thermal conductivity, and increases the local temperatures.

\section{Discussions}

As mentioned above, the NanoBTE solver supports both CPU- and GPU-accelerated implementations. After spectral and angular discretization, each propagation direction $\alpha$ and phonon band $\lambda$ defines an independent microscopic transport equation on the same spatial mesh. In the implementation, all $(\alpha,\lambda)$ pairs are collected as independent transport tasks and distributed among MPI ranks in a round-robin manner. The full spatial mesh and phonon distribution arrays are replicated on each rank, while each rank solves only its assigned subset of band-direction transport equations. This strategy avoids spatial-domain decomposition and therefore simplifies the treatment of unstructured meshes, complex boundary conditions, and global temperature updates. The CPU implementation forms the numerical backbone of NanoBTE and is built on the scientific-Python stack of \texttt{NumPy}, \texttt{SciPy}, and \texttt{mpi4py}. The finite-volume discretization is assembled entirely with sparse linear algebra: for each propagation direction $\alpha$ a unit advection operator $K_\alpha$ is built as a \texttt{scipy.sparse} matrix (in coordinate format and converted to compressed-sparse-column storage), and the transport matrix of each band--direction task is obtained as $A_{\alpha\lambda}=\tau_\lambda^{-1}\mathbf{I}+v_\lambda K_\alpha$. A Fourier diffusion solve provides the initial temperature guess, whose Dirichlet matrix is factorized once with the sparse LU routine \texttt{scipy.sparse.linalg.splu} and reused through non-orthogonal correction iterations. Within each transport step, the correction $\Delta e_{\alpha\lambda}$ is obtained either from a cached \texttt{splu} factorization, reused across outer iterations since $A_{\alpha\lambda}$ is geometry- and mode-dependent but iteration-independent, or, optionally, from a matrix-free \texttt{bicgstab} iteration preconditioned by the Jacobi (diagonal) approximation via a \texttt{LinearOperator}, with a direct \texttt{spsolve} used as a robust fallback. The macroscopic couplings, namely the moment reductions for $T_L$, $T$, and $\mathbf{q}$, the vertex interpolation operator, and the owner-cell divergence of $\mathbf{q}_{\mathrm{nonF}}$, are expressed as vectorized \texttt{NumPy} operations and \texttt{bincount}-based scatter reductions, so that the synthetic macroscopic diffusion equation is likewise assembled sparsely and solved by a single cached LU factorization. Finally, because the $(\alpha,\lambda)$ tasks are distributed round-robin across ranks while the mesh and distribution arrays are replicated, the per-rank contributions are merged with a single \texttt{mpi4py} \texttt{Allreduce} at each iteration, which keeps the communication pattern simple and independent of the unstructured mesh partitioning. To evaluate its parallel-computing performance, we performed benchmark calculations on the three smallest benchmark cases considered in this study, including the two-dimensional in-plane MoS$_2$ case with $L = 10$ nm, the FinFET case with a fin width of 10 nm, and the GAAFET case with a nanosheet height of 5 nm. The CPU benchmarks were performed on a dual-socket Intel Xeon Gold 6133 platform. This platform provides 80 logical CPUs, corresponding to 40 physical cores with two hardware threads per core. Each socket contains 20 physical cores, and the system comprises two NUMA nodes.

\begin{figure}[ht!]
    \centering
    \includegraphics[width=0.95\linewidth]{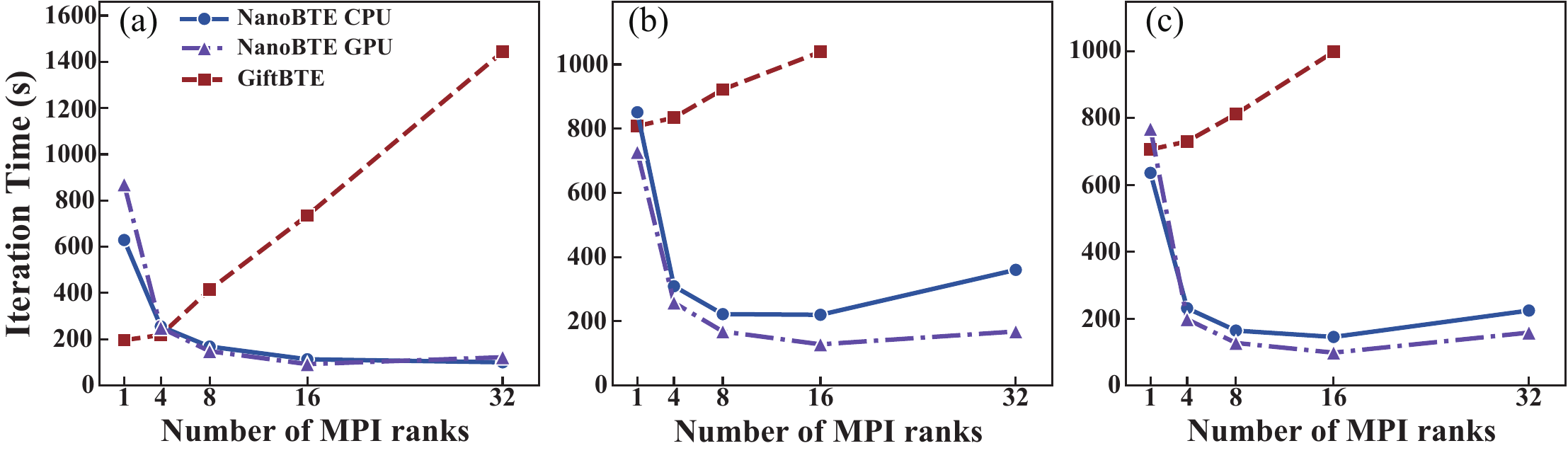}
    \caption{CPU and GPU performance comparison of NanoBTE with GiftBTE\cite{hu2024giftbte,hu2022optimized,hu2024ultraefficient} for the three smallest benchmark cases. The comparison includes NanoBTE CPU, NanoBTE GPU, and GiftBTE\cite{hu2024giftbte,hu2022optimized,hu2024ultraefficient} results. (a) The two-dimensional in-plane MoS$_2$ case with (L = 10 nm). (b) The FinFET case with a fin width of 10 nm. (c) The GAAFET case with a nanosheet height of 5 nm. The CPU benchmarks are performed on a dual-socket Intel Xeon Gold 6133 CPU platform. The GPU benchmarks are performed using two NVIDIA GeForce RTX 4090.}
    \label{fig:cpu}
\end{figure}

The iteration times plotted in Figure~\ref{fig:cpu} show that NanoBTE can effectively utilize CPU parallelism. For the in-plane MoS$_2$ case, the runtime decreases from 629 s on one core to 100 s on 32 cores. For the FinFET/GAAFET case, the runtime decreases from 851/636 s on one core to 220/145 s on 16 cores. In comparison, GiftBTE\cite{hu2024giftbte,hu2022optimized,hu2024ultraefficient} shows limited scalability in the present CPU benchmarks. Consequently, the relative advantage of NanoBTE becomes more pronounced under multi-core execution. The improved CPU scaling of NanoBTE is closely related to its task-parallel organization of the discretized phonon BTE in the band-direction phase space.

The GPU-accelerated version of NanoBTE is evaluated on a workstation equipped with two NVIDIA GeForce RTX 4090 GPUs, each providing 24~GB of device memory. The GPU implementation retains the same band--direction task decomposition as the CPU version: the $(\alpha,\lambda)$ transport tasks are distributed among MPI ranks in the same round-robin manner, and the ranks share the two available GPUs. To characterize the parallel behavior of this configuration, the GPU runtime is measured over the same range of MPI rank counts (1--32) used in the CPU benchmark, and the results are also reported in Figure~\ref{fig:cpu}.

As the number of MPI ranks increases from 1 to 16, the GPU runtime decreases rapidly for all three cases, reaching a minimum of 92~s for the in-plane MoS$_2$ case, 128~s for the 10~nm FinFET, and 98~s for the 5~nm GAAFET. This corresponds to roughly $6\times$ to $9\times$ acceleration relative to the single-rank GPU runtime, and reflects the increasing overlap of independent band-direction transport solves as more ranks concurrently issue GPU kernels. Beyond 16 ranks, however, the trend reverses, and the runtime increases to 122~s, 168~s, and 158~s at 32 ranks, respectively. This degradation arises because 32 MPI ranks share only two physical GPUs, so that each device is oversubscribed by 16 ranks; the resulting contention for GPU memory and kernel scheduling outweighs the benefit of additional concurrent tasks. For the present dual-GPU platform, the optimal configuration is therefore 16 ranks, at which each device is shared by a moderate number of ranks without severe contention.

As shown in Figure~\ref{fig:cpu}, since the CPU-NanoBTE and GPU runtimes are shown at identical MPI rank counts, the two back-ends can be compared directly under the same degree of task parallelism. In the intermediate range from 4 to 16 ranks, the GPU is consistently faster than the CPU for all three cases; at 16 ranks, for instance, the GPU reduces the runtime from 113 to 92~s for the in-plane case ($1.2\times$), from 220 to 128~s for the FinFET ($1.7\times$), and from 145 to 98~s for the GAAFET ($1.5\times$). At a single rank, by contrast, the GPU offers little advantage and is even slower than the CPU for the in-plane and GAAFET cases (870 versus 629~s and 768 versus 636~s), because the kernel-launch and host-device transfer overhead is not amortized when only one transport task is active per device. At 32 ranks, both back-ends lose efficiency, but for different reasons: the CPU runtime of the two three-dimensional cases degrades beyond its 16-core optimum (360~s for the FinFET and 224~s for the GAAFET) as the fixed problem is spread too thinly across cores, whereas the two shared GPUs become oversubscribed. The net outcome is that the GPU retains a clear advantage for the FinFET and GAAFET at 32 ranks, while the CPU regains the lead for the in-plane case (100 versus 122~s), for which its runtime continues to improve up to 32 cores. Overall, the same-rank comparison indicates that GPU acceleration is most beneficial for the larger three-dimensional device geometries, where the per-task sparse transport solves are heavy enough to keep the GPUs well utilized.

The GPU back-end is built on CuPy (using its \texttt{cupyx.scipy.sparse} module for sparse operators) and shares the same \texttt{mpi4py}-based band--direction task decomposition as the CPU version. Within each iteration the CPU retains the finite-volume bookkeeping, the sparse LU transport solves, and the macroscopic diffusion solve, whereas the GPU is dedicated to the dominant data-parallel reductions over the phonon phase space. Taking as input the non-equilibrium energy-density tensor $e_{\lambda i}$ that already resides in host memory, the GPU evaluates the moment reductions that yield the lattice temperature $T_L$, the macroscopic temperature $T$, and the heat-flux components $\mathbf{q}$, together with the divergence of the non-Fourier flux $\boldsymbol{\nabla}\cdot\mathbf{q}_{\mathrm{nonF}}$ required by the synthetic scheme, the latter being assembled through an owner-cell scatter reduction over active mesh faces. Only the resulting cell-wise fields, each of length $N_{\mathrm{cell}}$, are transferred back to the host, where they drive the next macroscopic temperature update and the convergence check. Because these returned arrays are orders of magnitude smaller than the full distribution tensor, the per-iteration host--device traffic is kept minimal, so that the GPU absorbs the heaviest reductions while the accuracy-critical sparse solves remain on the CPU.

To keep GPU memory usage bounded when multiple MPI ranks share the two GPUs, the transport kernels are executed in a chunked manner over spatial cells rather than transferring the full phonon distribution tensor to the GPU at once. Because each rank retains its own local copy of the mesh and distribution arrays, the aggregate device-memory footprint grows with the number of ranks assigned to each GPU; processing the cell dimension in chunks limits the peak temporary allocation and improves the robustness of the calculation for realistic device geometries. This streaming strategy is particularly important in the oversubscribed 32-rank regime, where many ranks compete for the same device memory.

\section{Conclusion}

In this work, we present NanoBTE, a deterministic finite-volume solver for the steady-state non-gray phonon Boltzmann transport equation, supporting complex 2D/3D geometries, band-resolved phonon properties, discrete-ordinate angular quadrature, and thermalizing, diffuse, and specular boundary conditions, and a band-direction task decomposition enables MPI-based CPU parallelization and GPU acceleration. 
Based on representative nanoscale transport benchmarks, NanoBTE reproduced the temperature and heat-flux distributions of the state-of-the-art BTE solver while substantially improving computational efficiency. These results establish NanoBTE as an accurate, efficient, and scalable framework for phonon Boltzmann transport simulations in realistic semiconductor nano devices.

% \section{Patents}

% This section is not mandatory, but may be added if there are patents resulting from the work reported in this manuscript.

\section{Declarations}

\subsection{Acknowledgments}

% Anyone who contributed towards the article but does not meet
% \href{http://www.icmje.org/recommendations/browse/roles-and-responsibilities/defining-the-role-of-authors-and-contributors.html}{the criteria}
% for authorship, including those who provided professional writing services or materials, should be acknowledged. Authors should obtain permission to acknowledge from all those mentioned in the Acknowledgments section. This section is not added if the author does not have anyone to acknowledge.

Not applicable.

\subsection{Authors’ contributions}

% Single author: \\
% The author contributed solely to the article.

% Two or more authors: \\
% Made substantial contributions to conception and design of the study and performed data analysis and interpretation: Salas H, Castaneda WV;

% Performed data acquisition, as well as provided administrative, technical, and material support: Castillo N, Young V

Chen HJ and Lin HX contributed equally to this work.\\
Conceived the project and secured funding: Zhang H;
Supervised the research: Zhang H, Liu JJ;
Implemented the CPU version code and analyzed the results: Chen HJ;
Implemented the GPU version code: Lin HX;
Performed the phonon transport calculations: Tian XL;
Wrote the original manuscript: Chen HJ, Lin HX, Tian XL;
Contributed to the discussion of the results and manuscript preparation: Chen HY, Zhang J, Yan SY, Shao HZ, Wu Y, Zhu HY, Zhang H

\subsection{Availability of data and materials}

% Authors should declare where the data supporting their findings can be found. Data can be deposited into data repositories or published as supplementary information in the journal. Authors who cannot share their data should state that the data will not be shared and explain it.

% If a manuscript does not involve such issue, please state “Not applicable.” in this section.

The source code of NanoBTE is publicly available on GitHub at
\url{https://github.com/fduai4s/nanoBTE}. The datasets generated and/or analyzed
during the current study are available from the corresponding author upon reasonable request.

\subsection{Financial support and sponsorship}

% If there are sources of funding for the study reported, any relevant grant numbers and the link of funder’s website should be provided if any. The role of the funding body in the experiment design, collection, analysis and interpretation of data, and writing of the manuscript should be declared: \\
% This work was supported by Grant name XX (No. XXXX; No. XXX)...

% If there is no grant: \\
% None.

This work was supported by the National Natural Science Foundation of China
(No. 12574243; No. 92580128), the National Key R\&D Program of China
(No. 2023YFA1608501), and the Shanghai Municipal Natural Science Foundation
(No. 24ZR1406600).

\subsection{Conflicts of interest}

% If there are any potential conflicts of interest that may be perceived as inappropriately influencing the representation or interpretation of reported research results, please declare here.

% If not, please write as “All authors declared that there are no conflicts of interest.”.

% Some authors may be bound by confidentiality agreements. In such cases, in place of itemized disclosures, we will require authors to state “All authors declare that they are bound by confidentiality agreements that prevent them from disclosing their conflicts of interest in this work.”.

% If authors are unsure whether conflicts of interest exist, please refer to the “Conflicts of Interest” of OAE
% \href{https://oaepublish.com/index.php/jsss/pages/view/editorial_policies}{Editorial Policies}
% for a full explanation.

All authors declared that there are no conflicts of interest.

\subsection{Ethical approval and consent to participate}

% Research involving human subjects, human material or human data must be performed in accordance with the
% \href{https://www.wma.net/policies-post/wma-declaration-of-helsinki-ethical-principles-for-medical-research-involving-human-subjects/}{Declaration of Helsinki}
% and approved by an appropriate ethics committee. An informed consent to participate in the study should also be obtained from participants, or their parents or legal guardians for children under 16. A statement detailing the name of the ethics committee (including the reference number where appropriate) and the informed consent obtained must appear in the manuscripts reporting such research.

% If the manuscript does not involve such issue, please state “Not applicable.” in this section.

Not applicable.

\subsection{Consent for publication}

% Manuscripts containing individual details, images or videos, must obtain consent for publication from that person, or in the case of children, their parents or legal guardians. If the person has died, consent for publication must be obtained from the next of kin of the participant. Manuscripts must include a statement that a written informed consent for publication was obtained. Authors do not have to submit such content accompanying the manuscript. However, these documents must be available if requested.

% If the manuscript does not involve this issue, state “Not applicable.” in this section.

Not applicable.

\subsection{Copyright}

© The Author(s) 2026.

\bibliographystyle{unsrt}

\bibliography{references}

\end{document}